\documentclass[aps,prl,twocolumn,footinbib,groupedaddress]{revtex4-2}

\usepackage{amsmath}
\usepackage{graphicx} 
\usepackage[breaklinks=true,colorlinks,citecolor=blue,linkcolor=blue,urlcolor=blue]{hyperref}
\usepackage{siunitx}
\usepackage{soul}
\usepackage{bm}
\usepackage[normalem]{ulem}
\usepackage{enumerate}  
\usepackage{float}
\usepackage{amssymb}
\usepackage{mathrsfs}
\newcommand{\ket}[1]{\vert #1 \rangle}
\newcommand{\bra}[1]{\langle #1 \vert}

\begin{document}

\title{Spin Grouping in Ring Cavity and Its Protection on Entangled States Transfer}

\author{Chang Li}

\affiliation {Key Laboratory of Atomic and Subatomic Structure and Quantum Control (Ministry of Education), Guangdong Basic Research Center of Excellence for Structure and Fundamental Interactions of Matter, School of Physics, South China Normal University, Guangzhou 510006, China} 

\affiliation {Guangdong Provincial Key Laboratory of Quantum Engineering and Quantum Materials, Guangdong-Hong Kong Joint Laboratory of Quantum Matter, South China Normal University, Guangzhou 510006, China}



\begin{abstract}
\textcolor{black}{
Long-range interactions are essential for large-scale quantum computation and quantum interconnections. 
Cavities provide a promising avenue to achieve long-range interaction by enhancing the coupling of remote qubits through shared cavity modes.
In this work, we investigate a spin array coupled to a ring cavity supporting two counterpropagating modes, focusing on the system's eigenstates and spin dynamics in the low-excitation regime.
We show that, under specific spatial configurations, the spins naturally self-organize into two groups, within which exciton transport is confined.
This spin-grouping mechanism preserves coherence between spins across the two groups, and is leveraged to deterministically transfer entangled states between remote spin pairs with additional dynamical addressing.
We further propose feasible implementations using atomic qubits or solid-state platforms.
Our scheme enables entangling remote spins within a cavity, highlighting potential applications in scalable quantum information processing.
}
\end{abstract}

\maketitle

\textit{Introduction}\textbf{---}
Long-range interactions and nonlocal entanglement are important for large-scale quantum computation and quantum simulation~\cite{RevModPhys.95.035002, defenu2024out, PRXQuantum.2.040101, richerme2014non, landig2016quantum, PhysRevA.72.052304}. 
In large-scale systems, qubits are usually spatially separated from each other. Long-range interaction enable effective connecting of distant qubits to efficiently implement specific quantum algorithms or quantum simulations requiring long-range interactions. Significant progress has been made, with systems containing hundreds of physical qubits demonstrating quantum advantage~\cite{arute2019quantum, google2020hartree, zhong2020quantum, madsen2022quantum, RevModPhys.95.035001, daley2022practical, shao2024antiferromagnetic, guo2024site, ebadi2022quantum}. To enhance connectivity, various protocols have been developed depending on the type of qubit system. For instance, neutral atoms and trapped ions can be physically shuttled into interaction regions to enable any-to-any coupling~\cite{pino2021demonstration, bluvstein2022quantum, dhordjevic2021entanglement}, while scalable protocols for quantum state transport have been demonstrated in superconducting quantum circuits~\cite{kurpiers2018deterministic, zhong2021deterministic, storz2023loophole, xiang2024enhanced}. However, achieving long-range interactions remains challenging, as these approaches are often time-consuming and require ancillary qubits or additional control devices.

Cavity-mediated interactions offer a promising framework for achieving long-range coupling between distant qubits, a concept rooted in cavity quantum electrodynamics (cavity-QED)~\cite{RevModPhys.91.025005, mivehvar2021cavity}. In these systems, qubits coupled to a shared cavity interact via electromagnetic fields. When a qubit excites a cavity mode, the photon can propagate to another distant qubit, effectively inducing coupling without requiring physical proximity. 
Recent developments in cavity-QED technology have driven significant experimental progress in exploring quantum phases of matter~\cite{guo2021optical, helson2023density, sauerwein2023engineering, norcia2018cavity, PhysRevX.11.041046, dreon2022self, young2024observing} and advancing quantum information processing~\cite{sillanpaa2007coherent, mi2018coherent, niu2023low, PRXQuantum.5.020308, hartung2024quantum}. 
Therefore, it is timely to explore theoretical protocols that leverage cavities to enhance the connectivity of atomic systems with finite interactions or solid-state systems with local interactions. Such protocols could pave the way for constructing short-range quantum networks on scalable experimental platforms, ultimately advancing large-scale, fault-tolerant quantum computation~\cite{laracuente2022modeling, wilde2010nonlocal, chandra2024non}.

In this work, we study a simple yet intriguing system: a structured spin array strongly coupled to a ring cavity. Our focus is on the influence of spin structure on spin dynamics under low-excitation conditions and its application to long-distance quantum state transfer within the spin array. A ring cavity supports two counterpropagating cavity modes, with photons accumulating traveling phase and interacting with spins. To describe this system, we extend the Tavis-Cummings model by incorporating the two cavity modes and the position-dependent phase of the light field~\cite{PhysRevA.87.043817, PhysRevB.102.144202, PhysRevApplied.21.044028}. 
Our analysis reveals the formation of two degenerate polariton pairs under specific spin structures. 
\textcolor{black}{
Notably, spins are divided into two groups, and excitations are confined to propagate only within each group when the interval between spins is an odd multiple of $\lambda/4$, where $\lambda$ is the wavelength of the emitted photons from spins.
This configuration intrinsically preserves coherence between the two groups, enabling high-fidelity entangled state transfer between distant spin pairs. 
Population decay can be effectively suppressed by tuning the coupling strengths of individual spins to the cavity modes, allowing for robust and remote entanglement transfer from an initial spin pair to a target pair via an adiabatic addressing protocol. 
Our findings are experimentally feasible using programmable atomic or solid-state qubit systems coupled to ring cavities, providing a promising approach to realizing long-range interactions and advancing scalable quantum information processing.
}


\begin{figure}[t]
  \centering
  \includegraphics[width=8.6cm]{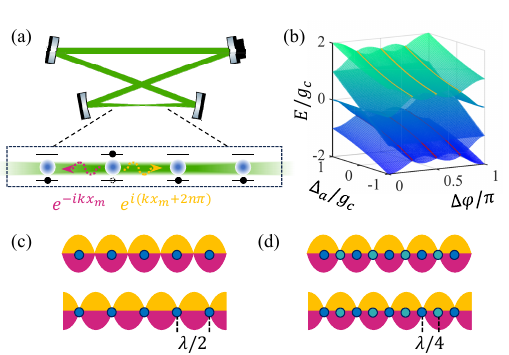}\\
  \caption{{A spin chain coupled to a ring cavity} (a) A schematic of four spins coupled to a bow-tie cavity. The spins are arranged with equal interval distances. The photon emitted from the second spin propagates through two possible cavity modes: the cw mode (yellow) and the ccw mode (fuchsia). As the photon travels through the cavity, it acquires a propagating phase, which depends on its position. (b) Energy levels of four spins coupled to the ring cavity. The eigenenergy, calculated in Eq.~\ref{eq2}, is plotted as a function of the interval phase $\Delta\varphi$ and spin detuning $\Delta_a$. The frequencies of the cavity modes and spins are set to $\omega_c=\omega_a=0$, with an additional detuning $\Delta_a$ applied to all the spins. The solid yellow and fuchsia lines represent the degeneracy of polaritons, and the degeneracy condition is solely dependent on $\Delta\varphi$, independent of $\Delta_a$. (c) Polariton eigenmodes with an interval distance of $\lambda/2$. The counterpropagating cw and ccw modes form nodes and antinodes, with spins aligned with the antinodes or nodes of the cavity modes corresponding to bright or dark polaritons, respectively. (d) Polariton eigenmodes with an interval distance of $\lambda/4$. The spins are divided into odd and even groups, with each group positioned at antinodes and coupled to the cavity modes, forming two individual pairs of polaritons.
	\label{fig1}}
\end{figure}


\textit{Extended Tavis-Cummings model \label{sectc}}\textbf{---}
We begin by considering a one-dimensional (1D) spin chain consisting of $N$ two-level spins trapped within a ring cavity. 
The spins are arranged at equal intervals and are homogeneously coupled to the cavity modes.
This bow-tie cavity supports two distinct cavity modes, propagating in the clockwise (cw) and counterclockwise (ccw) directions, respectively. 
The system is described by an extended Tavis-Cummings (TC) model with the Hamiltonian ($\hbar\equiv1$):
\begin{equation}
\label{eq1}
\begin{aligned}
\hat{H}=&\hat{H}_0+\hat{H}_I\\
=&\omega_c(\hat{a}^{\dagger}_{cw}\hat{a}_{cw}+\hat{a}^{\dagger}_{ccw}\hat{a}_{ccw})
+\omega_a\sum_{m=0}^{N-1}\hat{S}^{+}_m\hat{S}^{-}_m\\
&+\sum_{m=0}^{N-1}(g_{cw,m}\hat{a}^{\dagger}_{cw}\hat{S}_m^{-}+g_{ccw,m}\hat{a}^{\dagger}_{ccw}\hat{S}_m^{-})+H.C.,
\end{aligned}
\end{equation}
where $\hat{H}_0$ is the free Hamiltonian comprising the two cavity modes with equal cavity frequencies $\omega_c$ and N spins with identiacl energies $\omega_a$. Without loss of generality, we set $\omega_c=\omega_a=0$. The cavity modes are represented by the creation (annihilation) operators $\hat{a}^{\dagger}_{cw}$ ($\hat{a}^{}_{cw}$) and $\hat{a}^{\dagger}_{ccw}$ ($\hat{a}^{}_{ccw}$) as the cw and ccw modes, respectively. The $m$-th spin is described using the raising (lowering) operators, $\hat{S}_m^{+}$ ($\hat{S}_m^{-}$). The term $\hat{H}_I$ represents the TC coupling between the cavity modes and spins, under the rotating-wave approximation, with coupling strengths $g_{cw}$ and $g_{ccw}$. The emitted photon from a spin can travel alongside either cavity modes and accumulated propagating phase, as shown in Fig.~\ref{fig1}(a). The propagating phase is determined by the wave number of cavity mode, $k$, and the travelling distance of the photon. For each complete round trip in the cavity, the photon accumulates an integral multiple of $2\pi$. Consequently, the initial phase of an emitted photon is determined by the spin position, and the coupling terms are expressed as $g_{cw,m}=ge^{ikx_m}$ and $ g_{ccw,m}=ge^{-ikx_m}$, where $x_m$ denotes the position of m-th spin, and the coupling strengths are equally taken as $g$. 

In the following discussion, we restrict the single-excitation regime, where the total exciton number is conserved: $\hat{a}^{\dagger}_{cw}\hat{a}_{cw}+\hat{a}^{\dagger}_{ccw}\hat{a}_{ccw}+\sum_{m=0}^{N-1}\hat{S}^{+}_m\hat{S}^{-}_m \equiv1$. \textcolor{black}{Low-excitation condition provides an intuitive characterization on the eigenstates and eigenenergies of a many-body system under strong coupling~\cite{PhysRevB.109.064202}.}
The phase difference between adjacent spins is defined as $\Delta\varphi=k(x_{m+1}-x_m)$.
Under the single-excitation constraint, the Hamiltonian operates within a Hilbert space spanned by $N+2$ independent states: the photon states $\hat{a}^{\dagger}_{cw}\ket{vac}$, $\hat{a}^{\dagger}_{ccw}\ket{vac}$, and the spin excitation states $\hat{S}_m^{+}\ket{vac}$ ($m=0,...,N-1$), where $\ket{vac}$ represents the ground state with no excitations $\ket{0,0,g_0,...,g_{N-1}}$. 
The Hamiltonian in this single-excitation Hilbert space is then expressed as:
\begin{equation}
\label{eq2}
\begin{aligned}
\hat{H}=&\left[
\begin{matrix}
&H_c & G \\\
&G^{\dagger} & H_a
\end{matrix}
\right].
\end{aligned}
\end{equation}

Here, $H_c=\omega_c I_2$ represents the energy of the cavity modes, where $I_2$ is the identity matrix in the 2-dimensional subspace of cavity mode states. Similarly, $H_a=\omega_c I_N$ represents the spin energies, where $I_N$ is the identity matrix in the N-dimensional subspace of spin excitation states. The coupling between cavity modes and spins is given by:
\begin{equation}
\label{eq3}
\begin{aligned}
{G}=\frac{g_c}{\sqrt{N}}\left[
\begin{matrix}
&1,&...,& e^{im\Delta\varphi},&...,& e^{i(N-1)\Delta\varphi}\\\
&1,&...,& e^{-im\Delta\varphi},&...,& e^{-i(N-1)\Delta\varphi}
\end{matrix}
\right],
\end{aligned}
\end{equation}
where $g_c=g\sqrt{N}$ is the collective coupling strength of the spins to the cavity modes. By applying singular value decomposition (SVD), the coupling matrix $G$ can be represented as $G=U\Lambda W^{\dag}$,where $\Lambda$ is a diagonal matrix containing the singular values of $G$. These singular values represent the effective coupling strengths between the cavity modes and the spin chain.
Denoting the nonzero singular values as $\lambda_{\pm}=g_c\sqrt{1\pm |s|}$, we define $s$ as the structure factor of the spin chain, given by $s=\sum_{m=0}^{N-1}\exp{(2im\Delta\varphi)/N}$.
As a result, the Hamiltonian $\hat{H}$ features two pairs of polariton eigenstates. These eigenstates correspond to two bright modes of the spin chain that are individually coupled to the cavity modes, while the remaining $N-2$ dark states are decoupled from the cavity modes. Further details regarding the eigenstates and eigenenergies can be found in~\cite{SupM}.

In Fig.~\ref{fig1}(b), we present the energy level diagram for a system of 4 spins coupled to the cavity. When $\Delta\varphi=n\pi, n=0,1,\cdots$, the structure factor $s=1$, resulting in one of the coupling strengths $\lambda_{-}=0$, which causes its corresponding polariton pair to become dark. An illustrative scheme is shown in Fig.~\ref{fig1}(c). 
Although cw- and ccw modes are in the form of plane waves, together they can form a standing wave with a node spacing of half the wavelength, $\lambda/2$. Thus, a spin chain with a periodicity of $\lambda/2$ has two possible coupling scenarios with the cavity modes: the spin chain is either strongly coupled to the cavity when aligned with the antinode, or it is decoupled from the cavity modes when positioned at the node. 
Conversely, when $s=0$, the coupling strengths are equal $\lambda_{+}=\lambda_{-}$, and the two polariton pairs become degenerate. As the bold lines shown in Fig.~\ref{fig1}(b), there are total $N-1$ degenerate points, and the interval phase satisfies $\Delta\varphi=j\phi/N, j=1,...,N-1$. 

If the phase difference between adjacent spins becomes an odd multiple of $\pi/2$, spins can be divided into two groups, with each group coupling to one of the mixed cavity modes. After performing the SVD on the coupling matrix $G$, the coupling strengths are obtained from non-zero diagonal term of $\Lambda$, and the collective cavity and spin states are obtained from the eigen-states of $U$ and $W$. 
When the phase difference between the spins is $\pi/2$, for an even number of spins, the two pairs of polaritons become degenerate. In this case, the states can be written as:
\begin{equation}
\label{eq4}
\begin{aligned}
\ket{C_+}&=(\hat{a}_{cw}^{\dagger}+\hat{a}_{ccw}^{\dagger})\ket{vac}\\
\ket{C_-}&=(\hat{a}_{cw}^{\dagger}-\hat{a}_{ccw}^{\dagger})\ket{vac}\\
\ket{A_+}&=\sum_{l=0}^{N/2-1}(-1)^l\hat{S}_{2l}^{+}\ket{vac}\\
\ket{A_-}&=\sum_{l=0}^{N/2-1}(-1)^l\hat{S}_{2l+1}^{+}\ket{vac},
\end{aligned}
\end{equation}
where $\ket{C_i}$ and $\ket{A_i} (i=1,2)$ represent the corresponding cavity and spin modes of the polaritons. From Eq.~\ref{eq4}, we can clearly observe that the spins are divided into two groups based on their indices: one group consists of spins with even indices and the other with odd indices. Each group is individually coupled to a distinct cavity mode, as illustrated in Fig.~\ref{fig1}(d). 

\textcolor{black}{
Although the analysis above is based on the simplified assumption of an evenly spaced, even-numbered spin array under single excitation, the spin grouping mechanism remains valid regardless of the total spin number or specific arrangement—as long as the two groups are separated by an odd multiple of $\lambda/4$. The interaction between spins is mediated by photons, which form a standing wave through interference between the two counterpropagating cavity modes. As a result, one group of spins positioned at the nodes of the standing wave becomes effectively decoupled from the other group, preserving the group separation even in the presence of multiple excitations. More details are provided in~\cite{SupM}.
}


\begin{figure}[t]
  \centering
  \includegraphics[width=8.6cm]{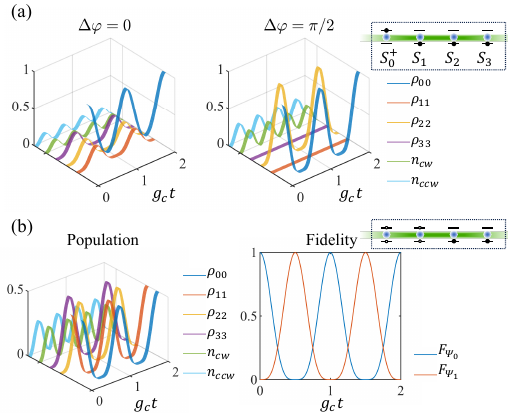}\\
  \caption{{Exciton transport} (a) Exciton transport under two structures. 
\textcolor{black}{The population of spins, $\rho_{ii}, (i=0,..,3)$ and photon in two cavity modes, $n_{cw}$ and $n_{acw}$, are shown as a function of time normalized with coupling strength, $g_c$, which are plotted as blue, orange, yellow, purple, green, cyan lines respectively.} The initial exciton is distributed in the first spin as indicated in the inset. When the interval phase $\Delta\varphi=0$, all spins are involved in the exicton oscillation. When $\Delta\varphi=\pi/2$, the exciton only oscillates between spin 0 and spin 2. (b) Entangled state transport. The exciton is initially equally distributed between spin pair (0, 1) as $\Psi_0$, and is then transferred to spin pair (2, 3) as $\Psi_1$. The population of spins and cavity modes and state fidelity are plotted as a function of time normalized with $g_c$.
	\label{fig2}}
\end{figure}


\textit{Exciton transport \label{sec3}}\textbf{---}
To confirm the spin grouping, we examine the exciton transport in a spin chain with different structure factors. \textcolor{black}{In the following sections, we numerically simulate the evolution of different initial states and analyze the resulting exciton distributions among the spins and cavity modes~\cite{SupM}}. As shown in Fig.~\ref{fig2}(a), we consider a 4-spin chain coupled to the ring cavity, with the first spin initially in the excited state, $\hat{S}_{0}^{+}\ket{vac}$. When structure factor $s=1$ and the conresponding phase difference $\Delta\varphi=0$, the exciton coherently oscillates across all cavity modes and spins. This is a trivial case since all spins are coupled to the cavity modes as part of a pair of bright polaritons.
In the case where the interval phase $\Delta\varphi=\pi/2$ and $s=0$, the exciton only oscillates within the cavity modes and the spin group with even indices. There is no population transfer to the odd-indexed spins, thus isolating the two groups under this structure. 
Therefore, we confirm the spin grouping by examining the exciton transfer for different structures, where the population in the two groups remains isolated when $\Delta\varphi=\pi/2$. Naturally, one might ask whether the coherence between the groups can be preserved. 

To address this, we consider an exciton equally distributed between two spins with an interval phase $\Delta\varphi=\pi/2$. This exciton can be represented as an entangled state: ${\Psi_0}=(\hat{S}_{0}^{+}+\hat{S}_{1}^{+})/\sqrt{2}\ket{vac}$. As shown in Fig.~\ref{fig2}(b), the exciton oscillates between spin pairs (0,1) and (2,3), mediated by the cavity modes. The oscillation frequencies of the two groups are identical because the eigenenergies of the two polaritons are degenerate. When the exciton is transferred to the spin pair (2,3), the spins become entangled, and the state can be written as ${\Psi_1}=(\hat{S}_{2}^{+}+\hat{S}_{3}^{+})/\sqrt{2}\ket{vac}$ with its fidelity reaching a maximum when the exciton is fully transported. Thus, spin grouping can protect the coherence during exciton transport, provided that entanglement is established between the two groups.


\begin{figure}[t]
  \centering
  \includegraphics[width=8.6cm]{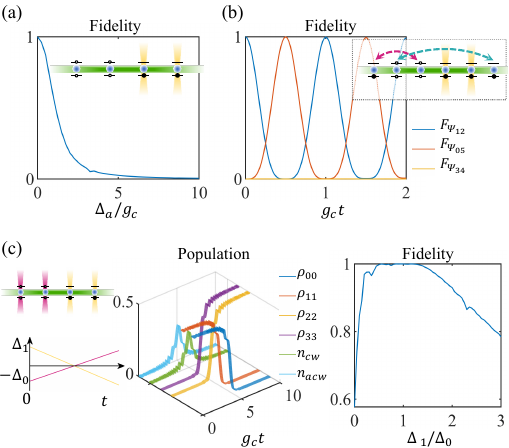}\\
  \caption{{Remote entangled state transfer.} (a) Spin pair (2, 3) are subject to additional detuning as shown in inset, and the maximum state fidelity of $\Psi_1$ is obtained within time $g_ct=10$. The fidelity decreases with larger detuning until it drops below $0.1\%$. (b) State fidelity as a function of time normalized with $g_c$. \textcolor{black}{Six spins are involved to demonstrate remote entangled state transfer as shown in the inset. The initial entangled state is $\Psi_{12}$, and additional detuning of $10g_c$ on spin pair (3, 4) prevents their participation in the state transfer process, therefore, the exciton in spin 1 (2) is transferred to spin 5 (0) as indicated by blue (red) dashed arrow.} The entangled state is then transferred to $\Psi_{05}$. 
(c) Entangled state transfer via STIRAP. \textcolor{black}{Both spin pair (0, 1) and (2, 3) are detuned by $\Delta_0$ (pink) and $\Delta_1$ (yellow) respectively with initial value $10g_c$. The detuning varies linearly with time, as shown below left. Population in spins and cavity modes is plotted as function of ramping time. The exciton is deterministically transferred from spin pair (0, 1) to (2, 3) after STIRAP}. Such process is robust against variations in $\Delta_1/\Delta_0$ from 0.5 to 1.4, with $\Delta_0=10g_c$ fixed. 
	\label{fig3}}
\end{figure}


\textit{Entangled state transfer \label{sec4}}\textbf{---}
In principle, an entangled state protected by spin grouping can be transferred to any position within the cavity, with the transfer speed determined solely by the coupling strength between the spins and cavity modes~\cite{SupM}. 
This capability is particularly advantageous for entangling remote qubits with finite interaction distances. For instance, it becomes feasible to entangle a single pair of spins with local control and subsequently transfer the entangled state to another spin pair positioned remotely. To facilitate this state transfer process, the detuning of specific spins should be dynamically adjusted, enabling them to couple to or decouple from the cavity modes as needed.

In Fig.\ref{fig3}(a), we show a scenario where four spins are coupled to the cavity modes with $\Delta\varphi=\pi/2$ with additional detuning applied to spins 2 and 3. We examine the maximum fidelity of the state ${\Psi_1}$ over 10 oscillation periods. As seen from the results, the fidelity rapidly drops below $1\%$ when the detuning $\Delta_a/g_c>7.5$. To further explore this, we consider a case with six spins coupled to the cavity, where two spins are detuned from the cavity modes with $\Delta_a/g_c=10$, as shown in Fig.~\ref{fig3}(b). The initial state is $\Psi_{12}=(\hat{S}_{1}^{+}+\hat{S}_{2}^{+})/\sqrt{2}\ket{vac}$ , and no population leakage occurs to the detuned spin pair (3,4). Meanwhile, the entangled state is transferred to spin pair (0,5) as $\Psi_{05}=(\hat{S}_{0}^{+}-\hat{S}_{5}^{+})/\sqrt{2}\ket{vac}$, achieving a maximum fidelity greater than $99.9\%$. Thus, applying local detuning to spins is confirmed as an effective method to control the entangled state transfer within the cavity.

The entangled state oscillates between the spin pairs coupled to the cavity, but it can be deterministically transferred using the STIRAP method~\cite{RevModPhys.89.015006}. As shown in Fig.~\ref{fig3}(c), we give an example where the initial state $\Psi_0$ is transferred to $\Psi_1$ via STIRAP process. The detuning of spins 0 and 1 is linearly ramped from $-\Delta_0$ to $\Delta_0$ while the detuning of spins 2 and 3 is ramped from $\Delta_1$ to $-\Delta_1$. As the detunings of two spin pairs are ramped, the entangled state $\Psi_0$ is gradually transferred to $\Psi_1$,  with a fidelity exceeding $99.9\%$ when $\Delta_0=\Delta_1=10g_c$. Even when $\Delta_1\neq \Delta_0$, the final state fidelity remains above $99.5\%$ for a ratio of $\Delta_1/\Delta_0$ ranging from $0.5$ to $1.4$, indicating that the entangled state transfer via the STIRAP method is robust against variations in detuning.


\begin{figure}[t]
  \centering
  \includegraphics[width=8.6cm]{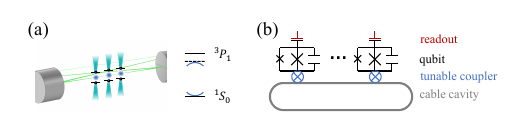}\\
  \caption{{Practical implementation} (a) Atomic array in tweezer traps coupled to a bow-tie cavity. $^{171}$Yb atoms are trapped in optical tweezers. The transition from $^{1}S_0$ to $^{3}P_1$ is coupled to the cavity modes, and an additional set of tweezers can be used to control the on-site detuning $\Delta_a$. (b) Superconducting qubits coupled to a cable cavity. The qubit frequencies and coupling strengths to the cavity modes are individually tunable to ensure uniform detuning and coupling for all qubits.
	\label{fig4}}
\end{figure}


\textit{Discussion}\textbf{---}
In the analysis above, we considered the ideal case, neglecting the spontaneous decay of the spins and the dissipation of the cavity modes. To implement our scheme practically, it is crucial to select an appropriate spin carrier such that the coupling strength is much larger than the dissipation rates of both the spins and cavity modes.

One promising candidate is the $^{171}$Yb atom, using the ground state $^{1}S_0$ and metastable state $^{3}P_1$, as shown in Fig.~\ref{fig4}(a). The spontaneous decay rate for this transition is $\Gamma=2\pi\times0.18$~MHz, and the cavity is supposed to have linewidth $\kappa=2\pi\times30$~KHz and cooperativity $C=200$. This gives a coupling strength $g=\sqrt{C\Gamma\kappa}/2=0.52$~MHz. For a system of four atoms coupled to the cavity, the collective coupling strength is approximately 1~MHz. However, the primary challenge remains the dissipation, which leads to exciton decay and thus reduces the fidelity of entangled states at the same rate~\cite{PhysRevResearch.6.L042026, chen2022high}.

Another candidate for spin carriers are superconducting qubits coupled to a cavity, which can achieve coupling strengths of up to 10~MHz, as shown in Fig.~\ref{fig4}(b). The decay rate for superconducting qubits is approximately 10~kHz, and the cavity dissipation rate is also around 10~kHz. In superconducting systems, the ratio between coupling strength and dissipation is much higher, making them a more promising candidate~\cite{PhysRevLett.127.107701, PRXQuantum.4.030336, song2024realization, deng2024long}. However, a technical challenge arises in ensuring that the resonance frequency of all qubits and their coupling strength to the cavity are identical.

\textcolor{black}{
In summary, we have analyzed a cavity-QED system composed of a spin chain coupled to a ring cavity, extending the TC model to incorporate two counterpropagating cavity modes and the associated photon traveling phase. 
The spin grouping mechanism preserves coherence between spin pairs, enabling high-fidelity entangled state transfer through adiabatic addressing. Our work contributes to the realization of long-range entanglement in cavity-QED systems without relying on direct spin-spin interactions, complementing other efforts such as cavity curving and cavity-mediated gate approaches~\cite{PhysRevLett.118.210503, grinkemeyer2025error}. 
Our scheme provides a practical method to enhance the connectivity of individual quantum processors and to interconnect multiple scalable processors.
Looking forward, it would be interesting to explore the interplay of photon and spin quantum phases in Dicke regime under strong external driving~\cite{PhysRevX.7.031002, PhysRevLett.125.257604, yang2021realization, PhysRevLett.129.253601, PhysRevLett.131.253603, baghdad2023spectral, PhysRevX.14.011026, PRXQuantum.6.020303}, moving beyond the low-excitation framework considered here.
}


\begin{acknowledgments}
We acknowledge insightful discussions with Zhi Li, Weilin Li and Wengang Zhang, and thank Hui Yan and Shiliang Zhu for helpful feedback on our manuscript. We acknowledge the use of the Quantum Toolbox in PYTHON (QuTiP)~\cite{JOHANSSON20131234}. This work is supported by the National Key R\&D Program of China (Grants N0.~2020YFA0309500 and No.~2022YFA1405300), the National Natural Science Foundation of China (Grants No. 12404407 and No.~12225405). All the data of figures in this work are available~\cite{li202414575410}.
\end{acknowledgments}

\bibliography{ref}

\newpage
\begin{widetext}
\title{Supplemental Material for \\ Spin Grouping in Ring Cavity and its Protection on Entangled States Transfer}

\setcounter{equation}{0}
\setcounter{figure}{0}
\setcounter{table}{0}
\setcounter{page}{1}
\setcounter{section}{0}
\makeatletter
\renewcommand{\theequation}{S\arabic{equation}}
\renewcommand{\thefigure}{S\arabic{figure}}


\maketitle

\section{Supplemental Material for \\ Spin Grouping in Ring Cavity and its Protection on Entangled States Transfer}

\section{I. Eigenmodes of Extended TC model \label{secAA}}

In this work, we extend the Tavis-Cummings model by considering two counterpropagating cavity modes with a traveling phase. The eigenenergies can be directly obtained from Eq.~2, as we only consider the single-excitation case. However, determining the spin grouping requires a more indirect approach. To achieve this, we apply the singular value decomposition (SVD) method on the coupling matrix $G$ from Eq.~3 to extract both the eigenenergies and eigenstates.

The system has $2+N$ eigenstates in the single-excitation manifold. These eigenstates include the single-photon states in the cavity modes ($\ket{C}_{cw}, \ket{C}_{ccw}$) and the excitations in the individual spins ($\ket{A}_m, m=1,...,N-1$), which are written as:
\begin{equation}
\label{eqa1}
\begin{aligned}
&\ket{C}_{cw}=\hat{a}^{\dag}_{cw}\ket{vac},\\
&\ket{C}_{ccw}=\hat{a}^{\dag}_{ccw}\ket{vac},\\
&\ket{A}_m=\hat{S}^{+}_m\ket{vac},
\end{aligned}
\end{equation}
where $\ket{vac}$ is the ground state of the system. Therefore the Hamiltonian in Eq.~1 can be written as:
\begin{equation}
\label{eqa2}
\begin{aligned}
\hat{H}=&\hat{H}_c+\hat{H}_a+\hat{H}_v\\
=&\omega_c\left(\ket{C_{cw}}\bra{C_{cw}}+\ket{C_{ccw}}\bra{C_{ccw}}\right)\\
+&\omega_a\sum_{m=0}^{N-1}\ket{A_m}\bra{A_m}\\
+&\sum_{m=0}^{N-1}\left(g_{cw,m}\ket{C_{cw}}\bra{A_m}+g_{ccw,m}\ket{C_{ccw}}\bra{A_m}\right)+H.C.
\end{aligned}
\end{equation}
The Hamiltonian can be further simplified as Eq.~2 and the coulping part $\hat{H}_v$ can be represented by $G$ as shown in Eq.~3. The coupling term, $G$, is a $2\times N$ matrix and can be decomposed as
\begin{equation}
\label{eqa3}
G=U\Lambda W^{\dag},
\end{equation}
where $U$ is a $2\times2$ unitary matrix, $\Lambda$ is an $2\times N$ diagonal matrix and $W$ is an $N\times N$ unitary matrix.  
There are two nonzero diagonal elements of $\Lambda$, noted as $\lambda_{\pm}$, since the number of spins exceeds the number of cavity modes.
Matrix $U$ ($W$) can be represented by the eigenstates of $\Sigma_c=GG^{\dag}$ ($\Sigma_a=G^{\dag}G$), and the nonzero eigenenergies of $\Sigma_c$ and $\Sigma_a$ are both equal to $\lambda_{\pm}^2$. Consequently, the corresponding eigenstates of the cavity and spin modes are coupled to form two pairs of polaritons.
As for the remaining $N-2$ spin states, they are decoupled from the cavity modes and do not contribute to the interaction part of the Hamiltonian, $H_v$. Therefore, the interaction part of Hamiltonian in Eq.~{\ref{eqa2}} can be decomposed as
\begin{equation}
\label{eqa4}
\begin{aligned}
\hat{H}_v&=\sum_{k=cw,ccw}\sum_{m=0}^{N-1}\sum_{j=\pm}U_{kj}\lambda_jW^{\dag}_{jm}\ket{C_k}\bra{A_m}+H.C.\\
&=\sum_{j=\pm}\left(\sum_{k=cw,ccw}U_{kj}\ket{C_k}\right)\lambda_j\left(\sum_{m=0}^{N-1}W^{\dag}_{jm}\bra{A_m}\right)+H.C.\\
\end{aligned}
\end{equation}
Then we obtain the cavity and spin states of polaritons, written as
\begin{equation}
\label{eqa5}
\begin{aligned}
\ket{C_{\pm}}&=\sum_{k=cw,ccw}U_{kj}\ket{C_k},\\
\ket{A_{\pm}}&=\sum_{m=0}^{N-1}W^{}_{jm}\ket{A_m},
\end{aligned}
\end{equation}
and eigenenergies of polaritons are $\lambda_{\pm}$.

To obtain $\lambda_{\pm}$, we may calculate the eigenstates of $\Sigma_c$, which is 
\begin{equation}
\label{eqa6}
\begin{aligned}
\Sigma_c=GG^{\dag}&=\frac{g_c^2}{N}
\left[
\begin{matrix}
&N,&\sum_{m=0}^{N-1}e^{2im\Delta\varphi}\\
&\sum_{m=0}^{N-1}e^{-2im\Delta\varphi},&N
\end{matrix}
\right]\\
&=g_c^2
\left[
\begin{matrix}
&1,&s\\
&s^*,&1
\end{matrix}
\right].
\end{aligned}
\end{equation}
The eigenenergies of Eq.~\ref{eqa6} is $\sigma_{\pm}=g_c^2(1\pm|s|)$, so that eigenenergies of polaritons are $\lambda_{\pm}=\sqrt{\sigma_{\pm}}=g_c\sqrt{1\pm|s|}$. 

Spin grouping can also be directly observed from the eigenstates of $\Sigma_a$, which is 

\begin{equation}
\label{eqa7}
\begin{aligned}
\Sigma_a&=G^{\dag}G\\
&=\frac{g_c^2}{N}\left[
\begin{matrix}
&2 &e^{i\Delta\varphi}+e^{-i\Delta\varphi} &e^{2i\Delta\varphi}+e^{-2i\Delta\varphi} &\cdots &e^{i(N-1)\Delta\varphi}+e^{-i(N-1)\Delta\varphi}\\
&e^{i\Delta\varphi}+e^{-i\Delta\varphi} &2 &e^{i\Delta\varphi}+e^{-i\Delta\varphi} &\cdots &e^{i(N-2)\Delta\varphi}+e^{-i(N-2)\Delta\varphi}\\
&e^{2i\Delta\varphi}+e^{-2i\Delta\varphi} &e^{i\Delta\varphi}+e^{-i\Delta\varphi} &2 &\cdots &e^{i(N-3)\Delta\varphi}+e^{-i(N-3)\Delta\varphi}\\
&\vdots&\vdots&\vdots&\ddots&\vdots\\
&e^{i(N-1)\Delta\varphi}+e^{-i(N-1)\Delta\varphi} &e^{i(N-2)\Delta\varphi}+e^{-i(N-2)\Delta\varphi} &e^{i(N-3)\Delta\varphi}+e^{-i(N-3)\Delta\varphi} &\cdots &2
\end{matrix}
\right].
\end{aligned}
\end{equation}

The collective spin states can be defined as the eigenstates of $\Sigma_a$ with nonzero eigenenergy, which are
\begin{equation}
\label{eqa8}
\ket{A_{\pm}}=\sum_{m=0}^{N-1}\left(e^{im\Delta\varphi}s^*\pm e^{-im\Delta\varphi}|s|\right)\ket{A_m}.
\end{equation}

To be specific, when $\Delta\varphi=\pi/2$ ,spin grouping can be directly observed from $\Sigma_a$, which becomes
\begin{equation}
\label{eqa9}
\begin{aligned}
\Sigma_a&=G^{\dag}G\\
&=\frac{g_c^2}{N}\left[
\begin{matrix}
&2 &0 &-2 &\cdots &0\\
&0 &2 &0 &\cdots &-2\\
&-2 &0 &2 &\cdots &0\\
&\vdots&\vdots&\vdots&\ddots&\vdots\\
&0 &-2 &0 &\cdots &2
\end{matrix}
\right],
\end{aligned}
\end{equation}
where the spin number is even. The nonzero elements can be divided into two subspace with even or odd row/column index, and the corresponding eigenstates are composed of odd or even spins. The collective spin states are
\begin{equation}
\label{eqa10}
\begin{aligned}
\ket{A_+}&=\frac{1}{\sqrt{2N}}\sum_{l=0}^{N/2-1}(-1)^l\ket{A_{2l}}\\
\ket{A_-}&=\frac{1}{\sqrt{2N}}\sum_{l=0}^{N/2-1}(-1)^l\ket{A_{2l+1}}.
\end{aligned}
\end{equation}

\textcolor{black}{
\section{II. Flip-flop spin interaction mediated by travelling photon}
}
\textcolor{black}{
Since our primary focus is on spin dynamics, the system can be simplified to include spins interacting via flip-flop processes mediated by the exchange of traveling photons in the cavity~\cite{evans2018photon}. This leads to an effective spin-spin interaction described by:
}
\begin{equation}
\label{eqz1}
\begin{aligned}
\hat{H_c}&=\sum_{i,j}\left(g_{cw,i}^*\hat{S}_i^+\hat{a}_{cw}g_{cw,j}\hat{a}_{cw}^{\dag}\hat{S}_j^-+g_{ccw,i}^*\hat{S}_i^+\hat{a}_{ccw}g_{ccw,j}\hat{a}_{ccw}^{\dag}\hat{S}_j^-\right)+H.C.\\
&=g_c^2\sum_{i,j}\left[e^{-ik(x_i-x_j)}\hat{S}_i^+\hat{S}_j^-+e^{ik(x_i-x_j)}\hat{S}_i^+\hat{S}_j^-\right]+H.C.,
\end{aligned}
\end{equation}
\textcolor{black}{
where $i, j$ index are the spins involved in flip-flop interactions. From Eq.~\ref{eqz1}, it is evident that the interaction between two spins vanishes when their separation is an odd multiple of $\lambda/4$, consistent with the conclusions drawn in Section I. This behavior arises because the two counterpropagating cavity modes interfere to form a standing wave; spins located at the nodes of this standing wave become decoupled from the traveling photons, while those at the antinodes experience maximal coupling. Consequently, spin grouping can emerge beyond the idealized case of equally spaced, even-numbered spins, and is generally expected for any configuration in which spin separation satisfies the multiple of $\lambda/4$.
}

\section{III. Energy level diagram\label{secAB}}
There are two polaritons and addtional $N-2$ spin states decoupled with cavity modes, so that Hamitonian in Eq.~\ref{eqa2} can be re-writed as
\begin{equation}
\label{eqb1}
\begin{aligned}
\hat{H}&=\omega_c\sum_{j=\pm}\ket{C_j}\bra{C_j}+\omega_a\sum_{j=\pm}\ket{A_j}\bra{A_j}\\
&+\sum_{j=\pm}\lambda_j\ket{C_j}\bra{A_j}+H.C.\\
&+\omega_a\sum_{i\ne\pm}\ket{A_i}\bra{A_i}.
\end{aligned}
\end{equation}
Here we suppose cavity modes (and spins) have the same frequency $\omega_c$ $(\omega_a)$, and corresponding energy level diagram is shown in Fig.~1(b)where the spin detuning and interval phase are varied. In Fig.~\ref{figa1}, we present slices of the energy diagram for a clearer view of its details. When the interval phase  $\Delta\varphi=0$, one pair of polariton has zero eigenergy, $\lambda_-=0$, which means the cavity mode is decoupled with spins. Consequently, there is a zero-energy level that is independent of the spin detuning variation.
When $\Delta\varphi=\pi/2$, the two pairs of polaritons become degenerate, leading to three energy levels corresponding to the polaritons and spin states that are decoupled from the cavity modes. There are $N-1$ degenerate points in total, with interval phase satisfying $\Delta\varphi=l/N, (l=1,\cdots,N-1)$ and $s=0$. 
Although the discussion here focuses on the case of an even number of spins, spin grouping also exists for an odd number of spins, as $\Sigma_a$ still has grouped nonzero elements. Coherence between the odd and even groups remains, but the exciton transport speed differs between the two groups. This is because the polaritons are not degenerate at $\Delta\varphi=\pi/2$. Therefore, the method for entangled state transfer described in this work is not applicable to systems with an odd number of spins.

\begin{figure*}[h]
  \centering
  \includegraphics[width=8.6cm]{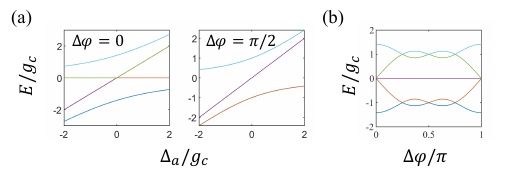}\\
  \caption{{Energy diagrams of four spins coupled to ring cavity} (a) Energy levels with varing spin detuning. (b) Energy levels with varying intervel phase when $\Delta_a=0$. 
	\label{figa1}}
\end{figure*}

\section{IV. State Transfer \label{secAC}}
We demonstrate the coherence between groups by showing the transfer of an entangled state between spin pairs, as illustrated in Fig.~2(b). In this section, we provide additional examples of exciton transport involving different cavity modes. 
In the first example, the initial entangled state is given by $\Psi_0=(\hat{S}_0^{\dag}+i\hat{S}_1^{\dag})/\sqrt{2}\ket{vac}$, and only cw-cavity mode is involved in transferring the state to $\Psi_1=(\hat{S}_2^{\dag}+i\hat{S}_3^{\dag})/\sqrt{2}\ket{vac}$, as shown in Fig.~\ref{figa2}(a). In the second example, if the initial state is $\Psi_0=(\hat{S}_0^{\dag}-i\hat{S}_1^{\dag})/\sqrt{2}\ket{vac}$, the ccw-mode participates in the state transfer, as shown in Fig.~\ref{figa2}(b). 

In both cases, the collective spin states and cavity modes are degenerate, and the relative phase of the entangled states between the two groups arises from the superposition of polaritons. As a result, the corresponding cavity modes are correlated differently. From another perspective, the direction of the emitted photon is determined by the relative phase of the spins, which is similar with the chiral quantum interconnect~\cite{PhysRevA.102.053720, kannan2023demand}.

\begin{figure*}[h]
  \centering
  \includegraphics[width=8.6cm]{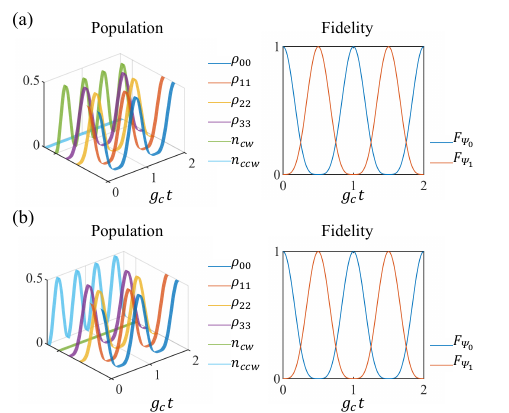}\\
  \caption{{Entangled State transfer} The population of spins and cavity modes and fidelity of entangled states are simulated, and the interval phase between adjacent spins is $\pi/2$. In (a), the initial state is $\Psi_0=(\hat{S}_0^{\dag}+i\hat{S}_1^{\dag})/\sqrt{2}\ket{vac}$ and target state is $\Psi_1=(\hat{S}_2^{\dag}+i\hat{S}_3^{\dag})/\sqrt{2}\ket{vac}$. In (b), the initial state is $\Psi_0=(\hat{S}_0^{\dag}-i\hat{S}_1^{\dag})/\sqrt{2}\ket{vac}$ and target state is $\Psi_1=(\hat{S}_2^{\dag}-i\hat{S}_3^{\dag})/\sqrt{2}\ket{vac}$.  
	\label{figa2}}
\end{figure*}
\textcolor{black}{
Not only the entangled state can be transferred with high fidelity, but also other states which are composed of spins from two groups. Here we show another two examples as shown in Fig.~\ref{figa4}. In the first example, the initial state is not a maximally entangled state, written as $\Psi_0=(0.6\hat{S}_0^{\dag}+0.8\hat{S}_1^{\dag})\ket{vac}$, population of which is not equally distributed in two spins. We can clearly observe in Fig.~\ref{figa4}(a) that population can only be transported within each spin group. Another example is that the initial state is $\Psi_0=\hat{S}_0^{\dag}\hat{S}_1^{\dag}\ket{vac}$, which contains two excitations as shown in Fig.~\ref{figa4}(b). This state can also be transferred to spin pair (2,~3) which is beyond single-excitation situation discussed in Section I.
}

\begin{figure*}[h]
  \centering
  \includegraphics[width=8.6cm]{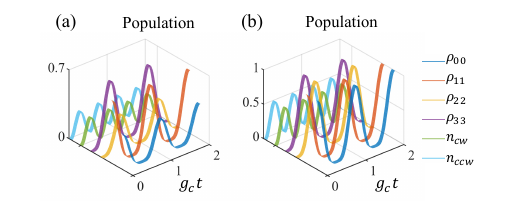}\\
  \caption{{State Transfer beyond maximal entanglement and single-excitation} The population of spins and cavity modes are simulated, and the interval phase between adjacent spins is $\pi/2$. In (a), the initial state is $\Psi_0=(0.6\hat{S}_0^{\dag}+0.8\hat{S}_1^{\dag})\ket{vac}$ and target state is $\Psi_1=(0.6\hat{S}_2^{\dag}+0.8\hat{S}_3^{\dag})\ket{vac}$. In (b), the initial state is $\Psi_0=\hat{S}_0^{\dag}\hat{S}_1^{\dag}\ket{vac}$ and target state is $\Psi_1=\hat{S}_2^{\dag}\hat{S}_3^{\dag}\ket{vac}$.  
	\label{figa4}}
\end{figure*}

\section{V. Fidelity of state transfer and STIRAP process \label{secAD}}
\textcolor{black}{
The dynamics of state transfer in this work is implemented using Quantum Toolbox in PYTHON (QuTiP). For N-spin case, we always set dimension of cavity mode to be $M=2N$, and obtain the state density matrix $\rho(t)$ after numerical simulating the time evolution of the system. THe population of spins, $\rho_{ii}$, and photon number in cavity modes, $n_{cw,ccw}$, are directly obtained from the state density matrix, and the state fidelity is calculated to quantify entanglement, written by:
\begin{equation}
\label{eqy1}
F_{\Psi_i}(t)=\bra{\Psi_i}\rho(t)\ket{\Psi_i},
\end{equation}
where $\Psi_i, (i=0,1)$, stands for the initial state or target states. 
}

Additional details about the STIRAP method used to transfer the entangled state are discussed here. For the numerical simulation, we first fix the STIRAP time to $g_ct=10$ and then determine the optimal detunings $\Delta_0$ and $\Delta_1$. We evaluate the final fidelity of the target state after the evolution, with an example shown in Fig.~\ref{figa3}(a). By scanning through different detuning values, we find that $\Delta_0=\Delta_1=10g_c$ provides one of the optimal conditions for achieving maximum fidelity of the target state, as shown in Fig.~\ref{figa3}(b).

\begin{figure*}[h]
  \centering
  \includegraphics[width=8.6cm]{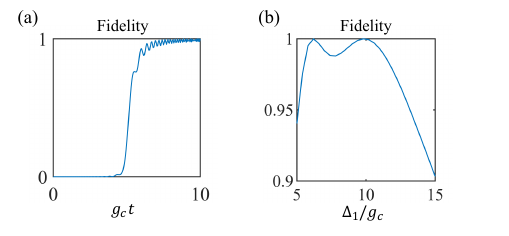}\\
  \caption{{State fidelity of STIRAP} (a) Target state fidelity evolution with $\Delta_0=\Delta_1=10g_c$. (b) Final state fidelity varying with spin detuning. Here we set $\Delta_0=\Delta_1$.  
	\label{figa3}}
\end{figure*}

\end{widetext}

\end{document}